# Demonstration of conditional gate operation using superconducting charge qubits

T. Yamamoto,[†‡] Yu. A. Pashkin,[‡*] O. Astafiev,[‡]

Y. Nakamura,[†‡] & J. S. Tsai[†‡]

[†]*NEC Fundamental Research Laboratories, Tsukuba, Ibaraki 305-8501, Japan*

[‡]*The Institute of Physical and Chemical Research (RIKEN), Wako, Saitama 351-0198, Japan*

[*] on leave from Lebedev Physical Institute, Moscow 117924, Russia

**Since the first demonstration of coherent control of a quantum state of a superconducting charge qubit[1] a variety of Josephson-junction-based qubits have been implemented[2-5] with remarkable progress in coherence time and read-out schemes. Although the current level of this solid-state device is still not as advanced as that of the most advanced microscopic-system-based qubits[6,7], these developments, together with the potential scalability, have renewed its position as a strong candidate as a building block for the quantum computer[8]. Recently, coherent oscillation[9] and microwave spectroscopy[10] in capacitively-coupled superconducting qubits have been reported. The next challenging step toward quantum computation is a realization of logic gates[11,12]. Here we demonstrate a conditional gate operation using a pair of coupled superconducting charge qubits. Using a pulse technique, we prepare different input states and show that they can be transformed by controlled-NOT (C-NOT) gate operation in the amplitude of the states. Although the phase evolution during the gate operation is still to be clarified, the present results are a major step toward the realization of a universal solid-state quantum gate.**

A Cooper-pair box provides an artificial two-level system where two charge states, say |0> and |1>, differing by 2*e* of one Cooper-pair (*e* is the electronic charge) are coherently superposed by Josephson coupling[13]. When two Cooper-pair boxes are connected by a capacitor, the quantum states of the boxes interfere with each other. This results in quantum beatings, as has been demonstrated recently[9]. Using this coherent four-level system formed by the charge states |00>, |01>, |10>, and |11>, we show how to implement a logic gate and demonstrate that it works as a quantum gate.

A scanning electron micrograph of the sample is shown in Fig. 1a. Two qubits are electrostatically coupled by an on-chip capacitor[9]. The right qubit has SQUID (superconducting quantum interference device) geometry and we use this qubit as the control qubit and the left one as the target qubit. Unlike the previous coupled-qubit sample[9] there are two independent pulse gates so that we can address each qubit individually. This is essential to the logic operation, as explained below.

In the two-qubit charge basis |00>, |10>, |01> and |11>, the hamiltonian of the system is given as

$$H = \sum_{n1,n2=0,1} E_{n1n2} |n1,n2\rangle\langle n1,n2| - \frac{E_{J1}}{2} \sum_{n2=0,1} (|0\rangle\langle 1| + |1\rangle\langle 0|) \otimes |n2\rangle\langle n2| \\ - \frac{E_{J2}}{2} \sum_{n1=0,1} |n1\rangle\langle n1| \otimes (|0\rangle\langle 1| + |1\rangle\langle 0|),$$  (1)

where $E_{J1}$ ($E_{J2}$) is the Josephson coupling energy of the first (second) box to the reservoir, $E_{n1n2} = E_{c1}(n_{g1}-n_1)^2 + E_{c2}(n_{g2}-n_2)^2 + E_m(n_{g1}-n_1)(n_{g2}-n_2)$ is the total electrostatic energy of the system ($n_1$, $n_2 = 0, 1$ is the number of excess Cooper pairs in the first and second boxes, and $n_{g1,2}$ are the gate-induced charges on the corresponding qubit divided by 2*e*). $E_{c1(2)} = 4e^2 C_{\Sigma 2(1)}/2(C_{\Sigma 1}C_{\Sigma 2} - C_m^2)$ are the effective Cooper-pair charging energies ($C_{\Sigma1(2)}$ are the sum of all capacitances connected to the corresponding island including the coupling capacitance $C_m$ between two boxes). Finally, $E_m = 4e^2 C_m/(C_{\Sigma1} C_{\Sigma2} - C_m^2)$ is the coupling energy. In our notation of $|n_1, n_2\rangle$ for the charge basis, $n_1$ and $n_2$ represent the states of the control and target qubits, respectively.

Figure 1b represents the idea for the gate operation. Using Eq. 1, we calculate the eigenenergies of the two-qubit system and plot them in the planes $n_{g1} = n_{g1}^0$ and $n_{g2} = n_{g2}^0$, where $n_{g1}^0$ and $n_{g2}^0$ are constants. In these planes, if $(n_{g1}^0, n_{g2}^0)$ is sufficiently far away from the co-resonant point[9] (0.5, 0.5), four energy bands can be regarded as two pairs of nearly independent single-qubit energy bands. In the plane of $n_{g1} = n_{g1}^0$, for example, our system is divided into a pair of independent two-level systems $|00\rangle$, $|01\rangle$ and $|10\rangle$, $|11\rangle$. Importantly, the charging energies of each of the two-level systems degenerate at different $n_{g2}$, namely, at $n_{g2L}$ for the states $|00\rangle$ and $|01\rangle$ and at $n_{g2U}$ for the states $|10\rangle$ and $|11\rangle$ as shown in Fig. 1b. This difference ($\delta n_{g2}$) originates from the electrostatic coupling between the qubits and is given as $E_m/2E_{c2}$. Similarly, we define $n_{g1L}$ and $n_{g1U}$ as shown in the plane of $n_{g2} = n_{g2}^0$.

Now we consider the pulse operation. Applying pulses to Pulse gate 1 (2) shifts the system non-adiabatically in the plane of $n_{g2} = n_{g2}^0$ ($n_{g1} = n_{g1}^0$). For convenience, we define the distances from $(n_{g1}^0, n_{g2}^0)$ to the degeneracy points as follows: $\delta n_{p1L} = n_{g1L} - n_{g1}^0$, $\delta n_{p1U} = n_{g1U} - n_{g1}^0$ and $\delta n_{p2L} = n_{g2L} - n_{g2}^0$. Suppose we start from the $|00\rangle$ state (point A) and apply an ideal rectangular pulse with an amplitude $V_{p2L} = 2e\, \delta n_{p2L}/C_{p2}$ to Pulse gate 2, where $C_{p2}$ is the capacitance between Pulse gate 2 and Box 2. This pulse is represented by the arrow in the ground-state charging diagram[14] of the base plane. In this case, the system is brought to the degeneracy point $n_{g2L}$ and evolves during a pulse duration $\Delta t$ with a frequency $\Omega = E_{J2}/\hbar$ between the |00> and the |01> states: $\cos(\Omega \Delta t/2)$ |00>+ $\sin(\Omega \Delta t/2)$ |01>. By adjusting $\Delta t$ so that $\Omega \Delta t = \pi$ ($\pi$ pulse), we can stop the evolution when the system is in the |01> state. The system is finally in the state at point C after the termination of the pulse.

On the other hand, if we start from the |10> state (point B) and apply the same pulse, the system does not reach the degeneracy point for states |10> and |11> ($n_{g2U}$). In this case, the amplitude of the oscillation between the |10> and the |11> states is suppressed by $E_{J2}^2/(E_m^2 + E_{J2}^2)$. If $E_m$ is sufficiently large, the state |10> remains almost unchanged (except for the phase factor), coming back to point B after the termination of the pulse. Similarly, we can realize the transition from the |01> state to the |00> state by the same pulse, and suppress the transition out of the |11> state. Therefore, conditional gate operation can be carried out based on this operation pulse: the target bit is flipped only when the control bit is |0>.

To experimentally demonstrate the above gate operation, we prepare different input states from the ground state |00> by applying pulses and measure the output of the gate operation. Figure 1c shows two pulse sequences that are utilized in the present experiment. For convenience, each of the pulses in the sequences is labelled by an index $m$ ($m$=1, …, 4, 5), which we will refer to as "Pulse $m$". In sequence (i) of Fig. 1c, a superposition of the states |00> and |10> is created by applying Pulse 1 with the amplitude $V_{p1L}=2e\, \delta n_{p1L}/C_{p1}$, where $C_{p1}$ is the capacitance between Pulse gate 1 and Box 1. In sequence (ii) of Fig. 1c, a superposition of the states |01> and |11> is created by two sequential pulses. First, Pulse 3, the same pulse as that for the gate operation, brings

the system to the |01> state at point C. Then, Pulse 4 with amplitude $V_{p1U}=2e\ \delta n_{p1U}/C_{p1}$ is applied.

In both sequences, an operation pulse (Pulse 2 or 5) creating an entangled state ($\alpha$|01>+$\beta$|10> or $\alpha$|00>+$\beta$|11>) is applied after the preparation pulses. To change the coefficients $\alpha$ and $\beta$, we change the Josephson energy of the control qubit $E_{J1}$ by a magnetic field, while keeping the pulse lengths constant. Because the control qubit has SQUID geometry, $E_{J1}$ is periodically modulated as $E_{J1}=E_{J1max}|\cos(\pi\ \phi_{ex}/\phi_0)|$, where $E_{J1max}$ is the maximum value of $E_{J1}$ and $\phi_0$ is the flux quantum. By repeatedly applying the sequential pulses (with a repetition time $T_r$=128 ns), we measure the pulse-induced currents through Probes 1 and 2, which are biased at ~650 µV to enable a Josephson-quasiparticle (JQP) cycle[15]. These currents are proportional to the probability of the respective qubit having one extra Cooper pair[1, 9].

Figure 2 shows the output currents of the control qubit ($I_C$) and the target qubit ($I_T$) as a function of $\phi_{ex}/\phi_0$ under the application of pulses shown in Fig. 1c (i). When no pulse is applied, both qubits show a finite current due to the finite width of the JQP peak (red curves in Fig. 2). Because this current depends on the Josephson energy, $I_C$ is periodically modulated by $\phi_{ex}$. First, we determine the length of the operation pulse (Pulse 2) by adjusting it to the peak in the single-qubit oscillation of $I_T$. When we apply Pulse 2 of this length (blue curves in Fig. 2), $I_T$ is enhanced and does not depend on $\phi_{ex}$, as was expected. Also, this pulse has no effect on $I_C$. Next, we apply the preparation pulse (Pulse 1) only. This pulse, in turn, induces current in $I_C$ while not affecting $I_T$ (green curves in Fig. 2). Furthermore, the magnitude of the induced current depends on $\phi_{ex}$, indicating that input states with different coefficients $\alpha$ and $\beta$ are prepared. Finally, we apply both Pulse 1 and Pulse 2 with an interval of 85 ps (orange curves in Fig. 2). In this case, $I_C$ shows the same dependence as that when only Pulse 1 is applied. However, $I_T$ also shows clear dependence on $\phi_{ex}$ and is anti-correlated with $I_C$ as the target qubit feels the state of the control qubit. In Fig. 3(a), we re-plot this data as a function of $E_{J1}$. We present only pulse-induced currents by subtracting the d.c. background currents from each curve. Both $I_T$ and $I_C$ show cosine-like dependence but their phases are opposite. That is, $I_T$ is maximal when $I_C$ is minimal, and vice versa. This is consistent with the expectation that the state $\alpha|01\rangle+\beta|10\rangle$ is created by the utilized pulse sequence.

Next we measure $\phi_{ex}$ dependence of $I_C$ and $I_T$ for pulse sequence (ii) of Fig. 1c (not shown) and plot it as $E_{J1}$ dependence in Fig. 3 (b). In this case, like in Fig. 3 (a), $I_T$ and $I_C$ show cosine-like dependence. However, most importantly, their correlation is now opposite to that in Fig. 3 (a). This is consistent with the expectation that the state $\alpha|00\rangle+\beta|11\rangle$ is created.

The above data shows that we have succeeded with the conditional gate operation. However, to understand more quantitatively, we compare the data with simulation data obtained by numerically calculating the time evolution of the density matrix. The results of the simulation are shown as black curves in Fig. 3. Here, we stress that no fitting parameters are used in the calculation.

First we consider the target qubit. Apart from the offset in Fig. 3(a), the simulated curves agree well with the experiment, suggesting that the oscillation amplitude of the measured $I_T$ is reasonable. On the other hand, we have some discrepancy in $I_C$. We attribute this discrepancy to the unknown current channel in our present read-out scheme. As long as the JQP process is considered, the pulse-induced current should not be able to exceed $2e/T_r$=2.5 pA, but in reality it does. This means that the pulse-induced current has an extra component that does not originate from the JQP process. We do not yet know the origin of this current. It may be other processes involving higher-order Cooper-pair tunnelling. The magnitude of this current probably depends on the Josephson energy (but does not depend strongly on the pulse length) and produces the $E_{J1}$-dependent deviation between the simulated and measured curves. In the target qubit, the similar current channel simply gives a constant offset in Fig. 3 as $E_{J2}$ is fixed and does not affect the overall $E_{J1}$-dependence. Although quantitative analysis for $I_C$ is difficult at present, the simulation suggests that the oscillation amplitude of the measured $I_T$ is reasonable, while that of $I_C$ is enhanced by this extrinsic factor originating from the imperfection of our read-out scheme.

Finally, we estimate the accuracy of our gate operation and propose possible ways for improvement. Our present read-out scheme, which does not allow us to measure the probability of the four states individually[9], makes it difficult to obtain the complete truth table of our gate operation solely from the experimental data. Instead, here we do it based on the simulation that turned out a reasonable description of our two-qubit system, as shown in Fig. 3. We calculate the time evolution of four perfect input states, |00>, |01>, |10> and |11> under the application of the operation pulse, namely Pulse 2 or 5 in Fig. 1(c) and plot the output probabilities as solid blue bars in Fig. 4. For the input states of |10> and |11>, our gate operation is almost ideal. Note that the accuracy is better than that expected for the case of the ideal pulse shape, that is 1- $E_{J2}^2$/ ($E_m^2+ E_{J2}^2$) ~ 0.84. This is due to the finite rise/fall time (40 ps) of the operation pulse, which suppresses the unwanted oscillation. On the other hand, for the input states of |00> and |01>, the output states have an unwanted component of |00> or |01> with a rather high probability.

This is also due to the finite rise/fall time, which in this case suppresses the desired oscillation. To improve this, increasing $E_m$ as well as making the pulse shape ideal would be the best solution. However, even with the present value of $E_m$, the simulation suggests that this matrix becomes much closer to the ideal one (keeping almost ideal outputs for |10> and |11> input states) if we slightly decrease the rise/fall time, say by 25% (red lines in Fig. 4), or decrease $E_{J2}$ by a similar amount.

In conclusion, we controlled our two-qubit solid-state circuit by applying a sequence of pulses and demonstrated the conditional gate operation. Although in the present experiment we paid attention only to the amplitude of the quantum state, phase evolution during the gate operation should also be examined for the realization of the quantum C-NOT gate (probably with additional phase factors), which is a constituent of the universal gate.

[1] Nakamura, Y, Pashkin, Yu. A. & Tsai, J. S. Coherent control of macroscopic quantum states in a single-Cooper-pair box. *Nature* **398**, 786-788 (1999).

[2] Vion, D *et al*. Manipulating the quantum state of an electrical circuit. *Science* **296**, 886-889 (2002).

[3] Yu, Y., Han, S., Chu, X., Chu, S.-I & Wang, Z. Coherent temporal oscillations of macroscopic quantum states in a Josephson junction. *Science* **296**, 889-892 (2002).

[4] Martinis, J. M., Nam, S., Aumentado, J. & Urbina, C. Rabi oscillations in a large Josephson-junction qubit. *Phys. Rev. Lett*. **89**, 117901-1 – 117901-4 (2002).

[5] Chiorescu, I., Nakamura, Y., Harmans, C. J. P. M. & Mooij, J. E. Coherent quantum dynamics of a superconducting flux qubit. *Science* **299**, 1869-1871 (2003).

[6] Vandersypen, L. M. K. *et al*. Experimental realization of Shor's quantum factoring algorithm using nuclear magnetic resonance. *Nature* **414**, 883-887 (2001).

[7] Gulde, S. *et al*. Implementation of the Deutsch-Jozsa algorithm on an ion-trap quantum computer. *Nature* **421**, 48-50 (2003).

[8] See for example, Nielsen, M. A. & Chuang, I. L. *Quantum Computation and Quantum Information.* (Cambridge Univ. Press, Cambridge, 2000).

[9] Pashkin, Yu. A. *et al*. Quantum oscillations in two coupled charge qubits. *Nature* **421**, 823-826 (2003).

[10] Berkley, A. J. *et al*. Entangled macroscopic quantum states in two superconducting qubits. *Science* **300**, 1548-1550 (2003).

[11] Shnirman, A., Schön, G. & Hermon Z. Quantum manipulations of small Josephson Junctions. *Phys. Rev. Lett*. **79**, 2371-2374 (1997).

[12] Averin, D. V. Adiabatic quantum computation with Cooper pairs. *Solid State Commun*. **105**, 659-664 (1998).

[13] Bouchiat, V., Vion, D., Joyez, P., Esteve, D. & Devoret, M. H. Quantum coherence with a single Cooper pair. *Physica Scripta* T**76**, 165-170 (1998).

[14] Pothier, H., Lafarge, P., Urbina, C., Esteve, D. & Devoret, M. H. Single-electron pump based on charging effects. *Europhys. Lett.* **17**, 249-254 (1992).

[15] Fulton, T. A., Gammel, P. L., Bishop, D. J., Dunkleberger, L. N. & Dolan, G. J. Observation of combined Josephson and charging effects in small tunnel junction circuits. *Phys. Rev. Lett*. **63**, 1307-1310 (1989).

[16] Nakamura, Y., Pashkin, Yu. A, Yamamoto, T. & Tsai, J. S. Charge echo in a Cooper-pair box. *Phys. Rev. Lett*. **88**, 047901-1 – 047901-4 (2002).

We thank B. L. Altshuler, D. V. Averin, S. Ishizaka, F. Nori, T. Tilma, C. Urbina, and J. Q. You for many fruitful discussions.

Correspondence should be addressed to T. Y. (e-mail: yamamoto@frl.cl.nec.co.jp).

Figure 1. Pulse operation of the coupled-qubit device. **a**, Scanning electron micrograph of the sample. The qubits were fabricated by electron-beam lithography and three-angle evaporation of Al on a $SiN_x$ insulating layer above a gold ground plane on the oxidized Si substrate. The two strips enclosed by red lines are the Cooper-pair boxes, which are coupled by an on-chip capacitor[9]. $\phi_{ex}$ represents magnetic flux penetrating the SQUID loop. An electrode between two pulse gates is connected to the ground to reduce the cross capacitance. Although there is a finite cross capacitance between one gate and the other box (about 15% of the main coupling), it does not play any essential role in the present experiment and so we can neglect it in this paper. The sample was cooled to 40 mK in a dilution refrigerator. The characteristic energies of this sample estimated from the d.c. current-voltage measurements are $E_{c1}$ = 580 μeV, $E_{c2}$ = 671 μeV and $E_m$ = 95 μeV. From the pulse measurements, $E_{J1}$ is found to be 45 μeV at a maximum and $E_{J2}$ to be 41 μeV. The superconducting energy gap is 209 μeV. Probe junction tunnel resistance is equal to 48 MΩ (left) and 33 MΩ (right). **b**, Energy band diagram along two lines of $n_{g1}=n_{g1}^0$ and $n_{g2}=n_{g2}^0$, where $n_{g1}^0$ and $n_{g2}^0$ are constants. Here $(n_{g1}^0, n_{g2}^0)$=(0.24,0.26), corresponding to the actual experimental condition. In the energy band diagram, black lines show the eigenenergies. The four coloured lines are the charging energies of the states shown in the cells of the charging diagram of the base plane with the corresponding colour. **c**, Pulse sequences used in the experiment. In both sequences, the upper and lower patterns show the pulse patterns applied to Pulse gates 1 and 2, respectively. The expected quantum

states after each pulse are also shown. The symbols |0> or |1> with subscripts $C$ and $T$ mean the state of the control and target qubits, respectively.

Figure 2. Magnetic-flux dependence of current of the control (top) and target (bottom) qubits under the application of pulses shown in Fig. 1c (i). The lengths of the pulses are $\Delta t_1$=85 ps, $\Delta t_2$=255 ps and $\Delta t_{12}$=85 ps, where we define the pulse length of Pulse $m$ in Fig. 1c as $\Delta t_m$ and the interval between pulses $l$ and $m$ as $\Delta t_{lm}$.

Figure 3. Pulse-induced current as a function of the Josephson energy of the control qubit under the pulse sequences shown in **a** Fig. 1c (i) and **b** Fig. 1c (ii). The lengths of the pulses in Fig. 1c (ii) are $\Delta t_3$=264 ps, $\Delta t_4$=88 ps, $\Delta t_5$=264 ps, $\Delta t_{34}$=88 ps and $\Delta t_{45}$=88 ps. The black curves represent the simulation obtained by calculating the time evolution of the density matrix. In the calculation, we assumed a trapezoidal pulse shape with both rise and fall times equal to 40 ps, which is close to the real pulse shape. To take into account the effect of dephasing, all the off-diagonal terms of the density matrix are set to zero before applying the operation pulse. This is a reasonable approximation because the dephasing time at an off-degeneracy point is reported to be a few hundred picoseconds[16], which is comparable to the time needed for the input preparation for the present experiment. We did not take into account the energy relaxation, which is known to be much slower.

Figure 4. Truth table of the present C-NOT operation estimated by the numerical calculation (solid blue bars). Detailed values of the probabilities

are $\begin{pmatrix} 0.37 & 0.62 & 0.004 & 0.003 \\ 0.62 & 0.37 & 0.004 & 0.007 \\ 0.004 & 0.004 & 0.97 & 0.018 \\ 0.003 & 0.007 & 0.018 & 0.97 \end{pmatrix}$. Ideally, they should be $\begin{pmatrix} 0 & 1 & 0 & 0 \\ 1 & 0 & 0 & 0 \\ 0 & 0 & 1 & 0 \\ 0 & 0 & 0 & 1 \end{pmatrix}$. We can partly see the correspondence of this figure to the experimental data in Fig. 3. Because the prepared input state in sequence (i) of Fig. 1c is almost pure |00> state when $E_{J1}$ equals zero, the $I_T$ at $E_{J1}$=0 in Fig. 3a normalized by the possible maximum current $2e/T_r$ (2.5 pA) should be close to 0.62 (the second element of the first column of the above truth table). The experimental data gives a slightly larger value ~0.8. This is attributed to the leak current discussed in the text. The red lines and arrows indicate the expected improvement after decreasing the rise/fall time of the pulses from 40 to 30 ps.

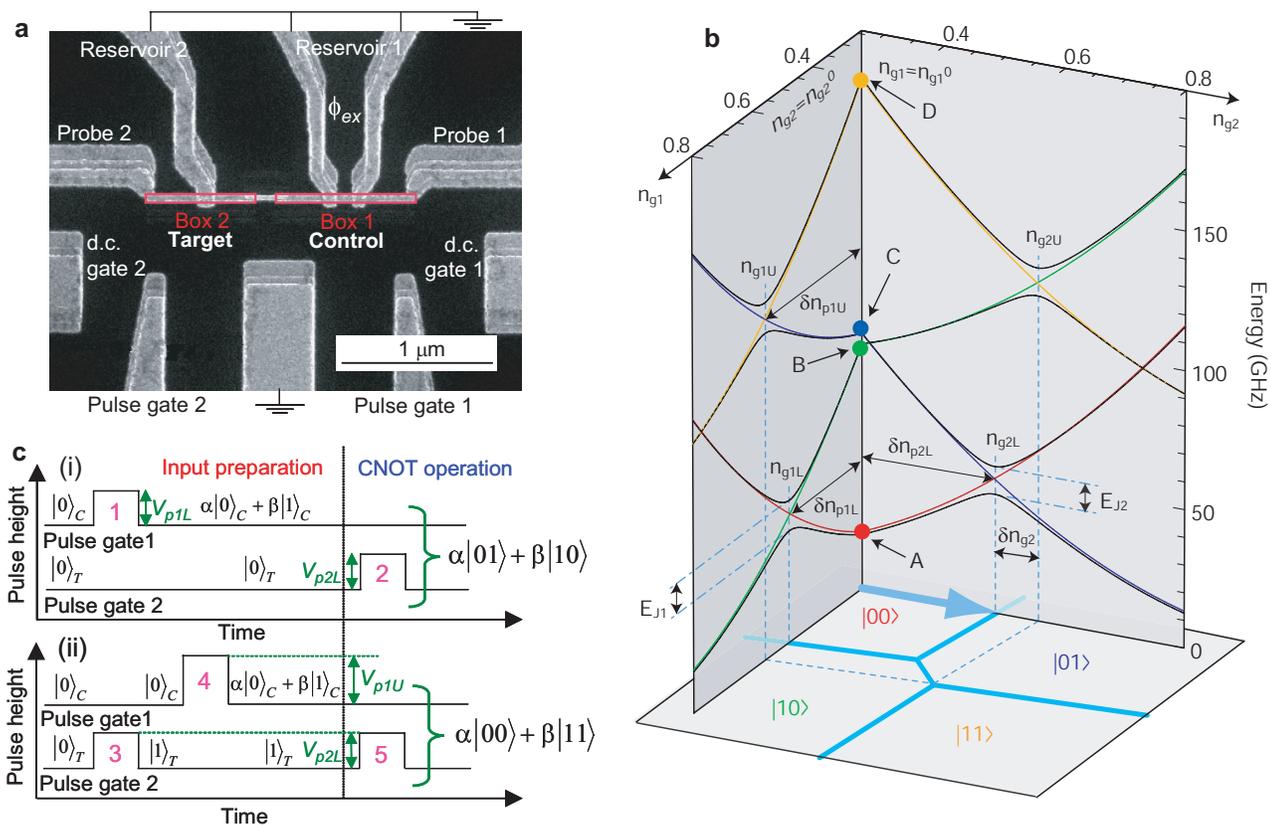

Yamamoto_fig.1

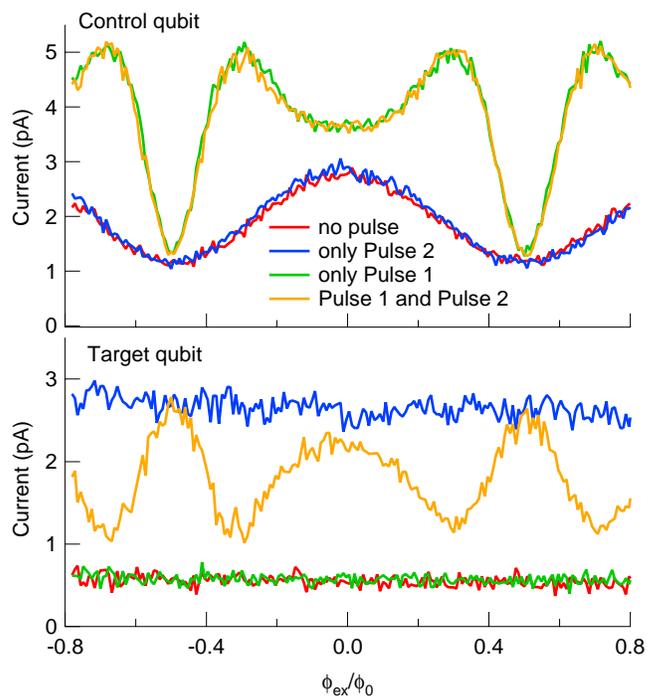

Yamamoto_fig.2

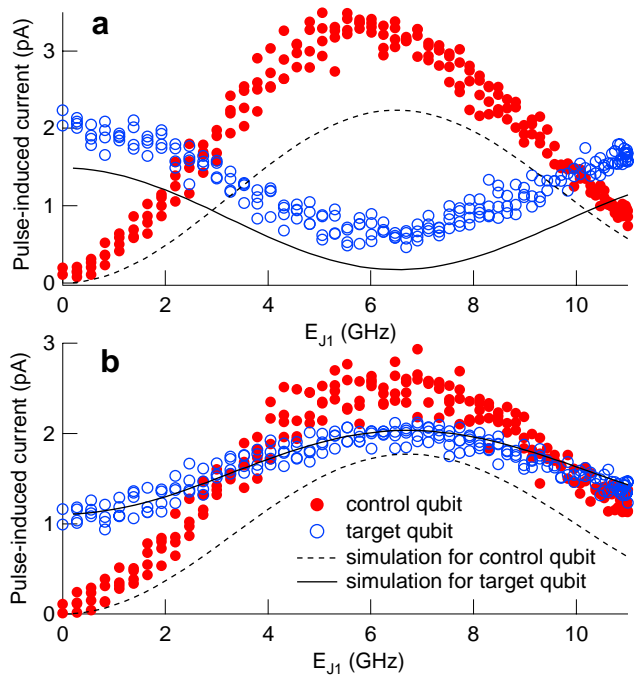

Yamamoto_fig.3

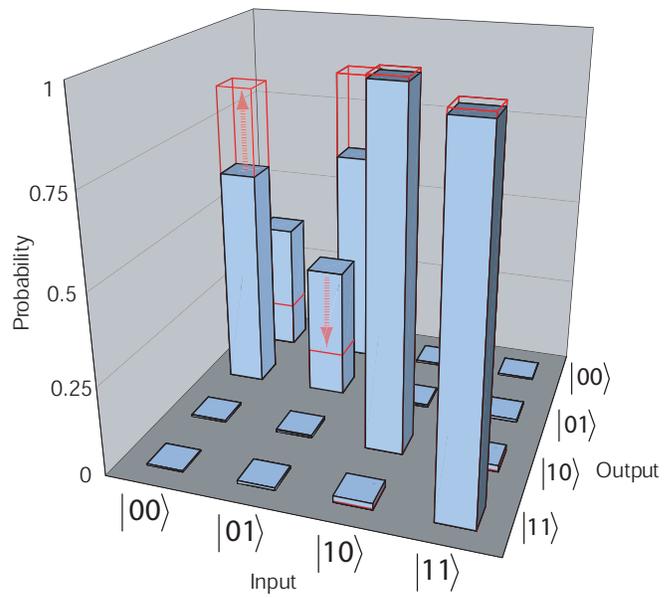

Yamamoto_fig.4